\documentclass[11pt,a4paper]{article}
\usepackage[T1]{fontenc}
\usepackage[latin9]{inputenc}
\usepackage{textcomp}
\usepackage{amsmath}
\usepackage{amssymb}
\usepackage{esint}

\makeatletter

\pdfpageheight\paperheight
\pdfpagewidth\paperwidth

\pdfoutput=1 

\usepackage{jheppub}

\usepackage{etoolbox}
    
    \patchcmd{\maketitle}{\@fpheader}{}{}{}

\usepackage[T1]{fontenc}
\usepackage{ae,aecompl}

\usepackage{amsfonts}

\usepackage{bbm}

\title{\boldmath Asymptotic structure of the Einstein-Maxwell theory on AdS$_{3}$}

\author[a]{Alfredo P\'{e}rez}
\author[a,b]{, Miguel Riquelme}
\author[a,c]{, David Tempo,}
\author[a]{and Ricardo Troncoso}

\affiliation[a]{Centro de Estudios Cient\'{i}ficos (CECs), Av. Arturo Prat 514, Valdivia,
Chile.}
\affiliation[b]{Departamento de F\'{i}sica, Universidad de Concepci\'{o}n, Casilla 160-C, Concepci\'{o}n, Chile.}
\affiliation[c]{Physique Théorique et Mathématique, Université Libre de Bruxelles and International Solvay Institutes, Campus Plaine C.P. 231, B-1050 Bruxelles, Belgium.}

\emailAdd{aperez@cecs.cl}
\emailAdd{riquelme@cecs.cl}
\emailAdd{tempo@cecs.cl}
\emailAdd{troncoso@cecs.cl}

\preprint{CECS-PHY-15/08}

\abstract{The asymptotic structure of AdS spacetimes in the context of General Relativity coupled to the Maxwell field in three spacetime dimensions is analyzed. Although the fall-off of the fields is relaxed with respect to that of Brown and Henneaux, the variation of the canonical generators associated to the asymptotic Killing vectors can be shown to be finite once required to span the Lie derivative of the fields. The corresponding surface integrals then acquire explicit contributions from the electromagnetic field, and become well-defined provided they fulfill  suitable integrability conditions, implying that the leading terms of the asymptotic form of the electromagnetic field are functionally related. Consequently, for a generic choice of boundary conditions, the asymptotic symmetries are broken down to $\mathbb{R}\otimes U\left(1\right)\otimes U\left(1\right)$. Nonetheless, requiring compatibility of the boundary conditions with one of the asymptotic Virasoro symmetries, singles out the set to be characterized by an arbitrary function of a single variable, whose precise form depends on the choice of the chiral copy. Remarkably, requiring the asymptotic symmetries to contain the full conformal group selects a very special set of boundary conditions that is labeled by a unique constant parameter, so that the algebra of the canonical generators is given by the direct sum of two copies of the Virasoro algebra with the standard central extension and $U\left(1\right)$. This special set of boundary conditions makes the energy spectrum of electrically charged rotating black holes to be well-behaved.}

\makeatother

\begin{document}
\maketitle \flushbottom

\section{Introduction}

The very first exact black hole solution endowed with a matter field
was the one found by Reissner and Nordström \cite{Reissner}, \cite{Nordstrom}.
The simple and remarkable properties that possesses have been widely
explored (see, e.g., \cite{Hawking-Ellis}, \cite{Townsed}, \cite{Romans}
and references therein), which appears to be a natural consequence
of the interaction of two of the greatest classical field theories
ever formulated: General Relativity and Electromagnetism. The Reissner-Nordström
solution can also be generalized to incorporate rotation \cite{Kerr},
\cite{Newman} even if the Einstein-Maxwell theory includes a cosmological
constant in diverse dimensions. However, in three-dimensional spacetimes,
the analysis of the properties of the corresponding electrically charged
(rotating) black hole solution \cite{BTZ}, \cite{Clement-Q+J}, \cite{MTZ}
can be somewhat darkened due to the logarithmic fall-off of the gauge
field, which severely modifies the asymptotic behaviour of the metric.
Indeed, finding a suitable definition of global conserved charges
in this context turns out to be a very subtle task \cite{Deser-Mazur},
\cite{MTZ}. Nonetheless, in the case of stationary spherically symmetric
solutions, it has been recently shown that the canonical generators
associated to the mass and the angular momentum can be naturally defined
without the need of any kind of regularization \cite{PRTT1}. Furthermore,
unlike the higher-dimensional case, the associated surface integrals
were found to acquire nontrivial contributions from the electromagnetic
field, and consequently, they become manifestly sensitive to the choice
of boundary conditions\footnote{We would like to emphasize that hereafter we make a distinction in
what we mean by ``asymptotic conditions'' and ``boundary conditions''.
For the former we refer to the fall-off of the fields in the asymptotic
region, which is an open set; while for the latter, we mean the conditions
that are held fixed at the boundary.}. This effect seems to reconcile the different results that have been
obtained for the mass of electrically charged black holes following
distinct regularization procedures \cite{ClementSpin}, \cite{Chan - Comment},
\cite{Kamata Koikawa mass}, \cite{MTZ}, \cite{Dias-Lemos}, \cite{Clement-Mass},
\cite{Cadoni-Mass}, \cite{Myung-Mass}, \cite{Jensen}, \cite{Garcia-review-mass},
\cite{Hendi-Mass}, \cite{Frassino =000026 Mureika}, since they might
correspond to inequivalent choices of boundary conditions. It is then
interesting to explore whether the effect extends to generic configurations,
which could be neither static nor spherically symmetric. With this
aim, in the next section we address some of the relevant geometric
aspects of asymptotically AdS$_{3}$ spacetimes in the Einstein-Maxwell
theory. They are shown to possess a relaxed behaviour compared with
the standard one of Brown and Henneaux \cite{Brown-Henne}. The canonical
realization of the asymptotic symmetries is then performed in section
\ref{sec:Canonical-generators}, where it is shown that the variation
of the generators associated to the asymptotic Killing vectors becomes
automatically finite once they are required to span the Lie derivative
of the fields. The surface integrals generically acquire contributions
from the electromagnetic field, and they are subject to nontrivial
integrability conditions that imply a functional relationship between
the leading terms of the asymptotic form of the electromagnetic field.
Since the canonical generators are shown to explicitly depend on the
choice of boundary conditions, in section \ref{sec:Compatibility}
we study the compatibility of different choices with the asymptotic
symmetries. It is shown that for generic boundary conditions the asymptotic
symmetries are broken down to $\mathbb{R}\otimes U\left(1\right)\otimes U\left(1\right)$.
However, there are two different sets of boundary conditions for which
only one of the chiral copies of the asymptotic Virasoro symmetries
is preserved. It is worth highlighting that requiring preservation
of the full conformal group singles out a very special set of boundary
conditions that is labeled by unique fixed constant parameter. The
canonical realization of the asymptotic symmetry algebra is then given
by the direct sum of $U\left(1\right)$ and two copies of the Virasoro
algebra with the standard central extension. Section \ref{Sec. Black Hole}
is devoted to compute the global charges of electrically charged rotating
black holes, which are found to be agree with the expressions that
were recently obtained in \cite{PRTT1} from the analysis of stationary
spherically symmetric configurations. We conclude with some additional
remarks in section \ref{Ending remarks}.

\section{Fall-off of the fields and asymptotic symmetries \label{sec:Geometric aspects}}

The asymptotic behaviour of the fields can be obtained following the
criteria described in \cite{Henne-Teitel AdS}, \cite{HMTZ-D}, \cite{HMT-Topologically},
\cite{HMT- More on...}. For a generic theory, the requirements to
be fulfilled can be spelled out as follows:
\begin{itemize}
\item The set has to contain as many asymptotic symmetries as possible.
\item The fall-off of the fields must be relaxed enough in order to accommodate
the solutions of physical interest.
\item Simultaneously, the decay has to be sufficiently fast so as to ensure
that the variation of the global charges is finite.
\item The boundary conditions must guarantee that the variation of the charges
can be integrated.
\end{itemize}
Once the four requirements aforementioned are taken into account for
asymptotically AdS$_{3}$ spacetimes in the Einstein-Maxwell theory,
one is led to propose the following fall-off for the metric and the
gauge field: 
\begin{align}
g_{\pm\pm} & =\frac{\kappa l^{2}}{4\pi^{2}}q_{\pm}^{2}\log\left(\frac{r}{l}\right)+f_{\pm\pm}+\mathcal{O}\left(\log\left(\frac{r}{l}\right)r^{-1}\right)\ ,\nonumber \\
g_{+-} & =-\frac{r^{2}}{2}+f_{+-}+\mathcal{O}\left(\log\left(\frac{r}{l}\right)r^{-1}\right)\ ,\nonumber \\
g_{rr} & =\frac{l^{2}}{r^{2}}+\frac{f_{rr}}{r^{4}}+\mathcal{O}\left(\log\left(\frac{r}{l}\right)r^{-5}\right)\ ,\label{eq:Asymptotic conditions metric}\\
g_{r\pm} & =\mathcal{O}\left(\log\left(\frac{r}{l}\right)r^{-3}\right)\ ,\nonumber \\
\nonumber \\
A_{\pm} & =-\frac{l}{2\pi}q_{\pm}\log\left(\frac{r}{l}\right)+\varphi_{\pm}+\mathcal{O}\left(\log\left(\frac{r}{l}\right)r^{-2}\right)\ ,\label{eq:Asymptotic conditions gauge field-2}\\
A_{r} & =\mathcal{O}\left(\log\left(\frac{r}{l}\right)r^{-3}\right)\ ,\nonumber 
\end{align}
where $f_{\pm\pm}$, $f_{+-},$ $f_{rr}$ and $q_{\pm}$ stand for
independent arbitrary functions of the null coordinates $x^{\pm}=\frac{t}{l}\pm\phi$.
It must be emphasized that the functions $\varphi_{\pm}$ turn out
to be functionally related with $q_{\pm}$ in a precise way. As explained
in section \ref{sec:Canonical-generators}, this has to be so in order
to ensure integrability of the variation of the global charges.

Note that the asymptotic behaviour of the metric in \eqref{eq:Asymptotic conditions metric}
might have included additional terms of the form $\mathcal{O}\left(\log\left(r/l\right)\right)$
and $\mathcal{O}\left(\log\left(r/l\right)r^{-4}\right)$, in $g_{+-}$
and $g_{rr}$ respectively, which we do not consider because they
can be consistently gauged away. Indeed, for this reason, the asymptotically
AdS$_{3}$ fall-off for the branch of stationary and spherically symmetric
spacetimes studied in \cite{PRTT1} fits within the asymptotic structure
described by \eqref{eq:Asymptotic conditions metric} and \eqref{eq:Asymptotic conditions gauge field-2}.
Consequently, the asymptotic behaviour proposed here not only accommodates
a wide set of exact solutions possessing these symmetries \cite{BTZ},
\cite{KK}, \cite{ClementSpin}, \cite{Hirschmann-Welch}, \cite{Cataldo-Salgado},
\cite{Cat-Salgado}, \cite{MTZ}, \cite{Cataldo}, \cite{Dias-Lemos},
\cite{crisostomo}, \cite{matyjasek}, \cite{Garcia annals}, but
also a larger class of configurations being neither stationary nor
spherically symmetric.

The asymptotic symmetries correspond to the subset of diffeomorphisms
and local $U\left(1\right)$ gauge symmetries, parametrized by $\xi^{\mu}=\xi^{\mu}\left(x^{\nu}\right)$
and $\eta=\eta\left(x^{\mu}\right)$ respectively, that preserve the
asymptotic behaviour of the fields, i.e., they have to fulfill
\begin{equation}
\begin{aligned}\delta_{\xi,\eta}g_{\mu\nu}=\pounds_{\xi}g_{\mu\nu} & =\mathcal{O}\left(g_{\mu\nu}\right)\ ,\\
\delta_{\xi,\eta}A_{\mu}=\pounds_{\xi}A_{\mu}+\partial_{\mu}\eta & =\mathcal{O}\left(A_{\mu}\right)\ .
\end{aligned}
\label{eq: Lie derivative metric-gauge potential}
\end{equation}
By virtue of \eqref{eq:Asymptotic conditions metric} and \eqref{eq:Asymptotic conditions gauge field-2}
the asymptotic Killing vectors and the asymptotic form of the $U\left(1\right)$
parameter are then found to be given by 
\begin{equation}
\begin{aligned}\xi^{\pm} & =T^{\pm}+\frac{l^{2}}{2r^{2}}\partial_{\mp}^{2}T^{\mp}+\mathcal{O}\left(\log\left(\frac{r}{l}\right)r^{-4}\right)\ ,\\
\xi^{r} & =-\frac{r}{2}\left(\partial_{+}T^{+}+\partial_{-}T^{-}\right)+\mathcal{O}\left(r^{-1}\right)\ ,\\
\eta & =\lambda+\mathcal{O}\left(\log\left(\frac{r}{l}\right)r^{-2}\right)\ ,
\end{aligned}
\label{eq:Asymptotic Killing vectors}
\end{equation}
 respectively, being described by three arbitrary functions $T^{\pm}=T^{\pm}\left(x^{\pm}\right)$
and $\lambda=\lambda\left(x^{+},x^{-}\right)$. Note that the subleading
terms of the asymptotic Killing vectors $\xi^{\pm}$ in \eqref{eq:Asymptotic Killing vectors}
decay slower than that of Brown and Henneaux \cite{Brown-Henne}. 

It must be highlighted that only a subset of the asymptotic symmetries
spanned by \eqref{eq:Asymptotic Killing vectors} are associated to
canonical generators. This is explained below.

\section{Canonical structure \label{sec:Canonical-generators}}

Let us consider the Einstein-Maxwell action in three spacetime dimensions,
given by

\begin{equation}
I\left[g_{\mu\nu},A_{\mu}\right]=\int d^{3}x\sqrt{-g}\left[\frac{1}{2\kappa}\left(R-2\Lambda\right)-\frac{1}{4}F_{\mu\nu}F^{\mu\nu}\right]\ ,\label{eq:Einstein-Maxwell action}
\end{equation}
where $\kappa$ and $\Lambda$ are related to the Newton constant
and the AdS radius according to $\kappa=8\pi G$ and $\Lambda=-l^{-2}$,
respectively. Assuming the standard ADM decomposition of the fields
(see, e.g., \cite{ADM}, \cite{Hanson Regge Teitelboim}), up to a
boundary term, the Hamiltonian form of the action reads 
\begin{equation}
I=\int d^{3}x\left\{ \pi^{ij}\dot{\gamma}_{ij}+p^{i}\dot{A_{i}}-N^{\perp}\mathcal{H}_{\perp}-N^{i}\mathcal{H}_{i}-\eta\mathcal{G}\right\} \ ,\label{eq:canonical hamiltonian action}
\end{equation}
where the constraints are given by
\begin{equation}
\begin{aligned}\mathcal{H}_{\perp} & =\frac{2\kappa}{\sqrt{\gamma}}\left(\pi^{ij}\pi_{ij}-\pi^{2}\right)-\frac{\sqrt{\gamma}}{2\kappa}\left(^{\left(2\right)}R-2\Lambda\right)+\frac{1}{2\sqrt{\gamma}}\gamma_{ij}p^{i}p^{j}+\frac{1}{4}\sqrt{\gamma}F_{ij}F^{ij}\ ,\\
\mathcal{H}_{i} & =-2\nabla_{j}\pi_{i}^{j}+p^{j}F_{ij}\ ,\\
\mathcal{G} & =-\partial_{i}p^{i}\ .
\end{aligned}
\end{equation}
Here $\pi^{ij}$ and $p^{i}$ stand for the momenta associated to
the dynamical fields $\gamma_{ij}$ and $A_{i}$, respectively. The
smeared total Hamiltonian then naturally splits as 
\begin{equation}
\begin{aligned}H\left(\xi,\eta\right) & =\int d^{3}x\left[\epsilon^{\perp}\mathcal{H}_{\perp}+\epsilon^{i}\mathcal{H}_{i}+\eta\mathcal{G}\right]\ ,\\
 & =\mathcal{H}\left(\xi\right)+\mathcal{G}\left(\eta\right)\ ,
\end{aligned}
\label{eq:Total Hamiltonian}
\end{equation}
with 
\begin{equation}
\begin{aligned}\epsilon^{\perp} & =N^{\perp}\xi^{t}\ ;\ \epsilon^{i}=\xi^{i}+N^{i}\xi^{t}\ .\end{aligned}
\end{equation}
It should be emphasized that the generator $\mathcal{H}\left(\xi\right)$
does not span the Lie derivative of the gauge field, but instead,
a ``pure diffeomorphism'' given by $\delta_{\xi}A_{\mu}=\left\{ A_{\mu},\mathcal{H}\left(\xi\right)\right\} =\xi^{\nu}F_{\nu\mu}.$
Hence, in order to make contact with the asymptotic symmetries in
\eqref{eq: Lie derivative metric-gauge potential}, the generator
has to be ``improved'' by the addition of a suitable $U\left(1\right)$
gauge transformation. The improved generator is obtained from \eqref{eq:Total Hamiltonian}
provided the parameter of the $U\left(1\right)$ gauge transformation
is of the form $\eta=\lambda+\xi^{\mu}A_{\mu}$ (see, e.g., \cite{Henneaux-Teitelboim  Z},
\cite{Bunster-Perez}), so that the total Hamiltonian now splits in
a different way, according to
\begin{equation}
H\left(\xi,\lambda\right)=\tilde{\mathcal{H}}\left(\xi\right)+\mathcal{G}\left(\lambda\right)\ ,
\end{equation}
 where 
\begin{equation}
\begin{aligned}\tilde{\mathcal{H}}\left(\xi\right) & =\mathcal{H}\left(\xi\right)+\mathcal{G}\left(\xi^{\mu}A_{\mu}\right)\ ,\end{aligned}
\label{eq:H magico}
\end{equation}
stands for the improved generator that fulfills what we were looking
for, i.e., 

\begin{equation}
\begin{aligned}\left\{ A_{\mu},H\left(\xi,\lambda\right)\right\} =\left\{ A_{\mu},\tilde{\mathcal{H}}\left(\xi\right)+\mathcal{G}\left(\lambda\right)\right\}  & =\text{£}_{\xi}A_{\mu}+\partial_{\mu}\lambda\ ,\end{aligned}
\end{equation}
on-shell. 

Remarkably, once the improved generator $\tilde{\mathcal{H}}\left(\xi\right)$
is supplemented by the appropriate boundary term that makes it to
be well-defined everywhere \cite{Regge-Teitelboim}, the variation
of the corresponding surface integrals becomes automatically finite.
Indeed, the surface integral associated to $\mathcal{G}\left(\xi^{\mu}A_{\mu}\right)$
precisely cancels out the divergences of the surface integral that
corresponds to the pure diffeomorphisms generator $\mathcal{H}\left(\xi\right)$.
It should be highlighted that this effect occurs due to the slow fall-off
of the electromagnetic field in three spacetime dimensions, so that
the improvement term amounts to an improper gauge transformation that
regularizes the global charges. Note that this is not the case for
asymptotically AdS spacetimes in $d\geq4$ dimensions, because the
improvement term in \eqref{eq:H magico} just corresponds to a proper
gauge transformation that does not modify the surface integrals associated
to the canonical generators. This is explicitly shown in what follows.

\subsection{Finite conserved charges and their integrability conditions }

The variation of the boundary terms which ensure that the canonical
generators $\tilde{\mathcal{H}}\left(\xi\right)$ and $\mathcal{G}\left(\lambda\right)$
are well-defined is denoted by $\delta Q_{\tilde{\mathcal{H}}}\left(\xi\right)$
and $\delta Q_{\mathcal{G}}\left(\lambda\right),$ respectively. By
virtue of the fall-off in \eqref{eq:Asymptotic conditions metric},
\eqref{eq:Asymptotic conditions gauge field-2}, the variation of
the $U\left(1\right)$ generator is finite and readily integrates
as 
\begin{equation}
Q_{\mathcal{G}}\left(\lambda\right)=\int dS_{l}\lambda p^{l}=\frac{1}{2\pi}\int d\phi\lambda\left(q_{+}+q_{-}\right)=\frac{1}{2\pi}\int d\phi\lambda q_{t}\ .\label{eq:electric charge}
\end{equation}

The variation of the surface integral associated to the improved generator,
given by $\delta Q_{\tilde{\mathcal{H}}}\left(\xi\right)$, is also
finite, but becomes subject to nontrivial integrability conditions. 

In order to see how the regularization mechanism is intrinsically
built in, following eq. \eqref{eq:H magico}, it is instructive to
split the variation of the surface integral according to

\begin{equation}
\begin{aligned}\delta Q_{\tilde{\mathcal{H}}}\left(\xi\right) & =\delta Q_{\mathcal{H}}\left(\xi\right)+\delta Q_{\mathcal{G}}\left(\xi^{\mu}A_{\mu}\right)\ ,\end{aligned}
\label{eq:Delta Q H tilde}
\end{equation}
 with 
\begin{equation}
\begin{aligned}\delta Q_{\mathcal{H}}\left(\xi\right) & =\int dS_{l}\left[\frac{1}{2\kappa}\epsilon^{\perp}G^{ijkl}\nabla_{k}\delta\gamma{}_{ij}-\frac{1}{2\kappa}\nabla_{k}\epsilon^{\perp}G^{ijkl}\delta\gamma{}_{ij}+2\epsilon^{j}\delta\left(\gamma_{jk}\pi^{kl}\right)\right.\\
 & \left.-\epsilon^{l}\pi^{jk}\delta\gamma_{jk}-\epsilon^{\perp}\sqrt{\gamma}F^{li}\delta A_{i}-\left(\epsilon^{l}p^{i}-\epsilon^{i}p^{l}\right)\delta A_{i}\right]\ ,\\
\delta Q_{\mathcal{G}}\left(\xi^{\mu}A_{\mu}\right) & =\int dS_{l}\xi^{\mu}A_{\mu}\delta p^{l}\ ,
\end{aligned}
\end{equation}
 where $G^{ijkl}=\frac{1}{2}\gamma^{1/2}\left(\gamma^{ik}\gamma^{jl}+\gamma^{il}\gamma^{jk}-2\gamma^{ij}\gamma^{kl}\right)$. 

Making use of the asymptotic fall-off of the fields \eqref{eq:Asymptotic conditions metric},
\eqref{eq:Asymptotic conditions gauge field-2}, as well as the asymptotic
Killing vectors \eqref{eq:Asymptotic Killing vectors}, the variation
of the surface integrals reduces to 
\begin{equation}
\begin{aligned}\delta Q_{\mathcal{G}}\left(T^{+},T^{-}\right) & =\delta Q_{\mathcal{G}}^{+}\left(T^{+}\right)+\delta Q_{\mathcal{G}}^{-}\left(T^{-}\right)\ ,\\
\delta Q_{\mathcal{H}}\left(T^{+},T^{-}\right) & =\delta Q_{\mathcal{H}}^{+}\left(T^{+}\right)+\delta Q_{\mathcal{H}}^{-}\left(T^{-}\right)\ ,
\end{aligned}
\end{equation}
with 
\begin{equation}
\begin{aligned}\delta Q_{\mathcal{\mathcal{G}}}^{\pm}\left(T^{\pm}\right) & =-\frac{1}{2\pi}\int d\phi T^{\pm}\left[\frac{l}{2\pi}\left(q_{\pm}\delta q_{+}+q_{\pm}\delta q_{-}\right)\log\left(\frac{r}{l}\right)-\varphi_{\pm}\left(\delta q_{+}+\delta q_{-}\right)\right]\ ,\\
\delta Q_{\mathcal{H}}^{\pm}\left(T^{\pm}\right) & =\frac{1}{2\pi}\int d\phi T^{\pm}\left[\frac{l}{2\pi}\left(q_{\pm}\delta q_{+}+q_{\pm}\delta q_{-}\right)\log\left(\frac{r}{l}\right)\pm q_{\pm}\left(\delta\varphi_{+}\mp\delta\varphi_{-}\right)\right]\\
 & +\frac{1}{2\pi}\int d\phi T^{\pm}\left[\delta\left(\frac{2\pi}{l\kappa}f_{\pm\pm}\mp\frac{l}{4\pi}q_{\pm}\left(q_{+}-q_{-}\right)\right)\right]\ .
\end{aligned}
\label{eq:Delta Q G+H}
\end{equation}

Therefore, it is clear that the logarithmic divergences in \eqref{eq:Delta Q G+H}
precisely cancel out, so that the variation of the improved generator
\eqref{eq:Delta Q H tilde} reads 
\begin{equation}
\begin{aligned}\delta Q_{\tilde{\mathcal{H}}}\left(T^{+},T^{-}\right) & =\delta Q_{\tilde{\mathcal{H}}}^{+}\left(T^{+}\right)+\delta Q_{\tilde{\mathcal{H}}}^{-}\left(T^{-}\right)=\int d\phi T^{+}\delta\mathcal{L}_{+}+\int d\phi T^{-}\delta\mathcal{L}_{-}\ ,\end{aligned}
\label{Delta Q H tilde T+-}
\end{equation}
where
\begin{equation}
\begin{aligned}\delta\mathcal{L}_{\pm} & =\frac{1}{2\pi}\delta\left(\frac{2\pi}{l\kappa}f_{\pm\pm}\mp\frac{l}{4\pi}q_{\pm}\left(q_{+}-q_{-}\right)\pm q_{\pm}\left(\varphi_{+}-\varphi_{-}\right)\right)+\frac{1}{2\pi}\left[\varphi_{+}\delta q_{-}+\varphi_{-}\delta q_{+}\right]\ .\end{aligned}
\label{DELTAL+-}
\end{equation}
Note that the terms within the square brackets in \eqref{DELTAL+-}
lead to nontrivial integrability conditions of the form
\begin{equation}
\begin{aligned}\delta^{2}Q_{\tilde{\mathcal{H}}}^{\pm} & =\end{aligned}
\frac{1}{2\pi}\int d\phi T^{\pm}\left[\delta\varphi_{+}\wedge\delta q_{-}+\delta\varphi_{-}\wedge\delta q_{+}\right]=0\ ,\label{eq:Integrability condition Q}
\end{equation}
which implies that $\varphi_{\pm}$ and $q_{\pm}$ are functionally
related. Assuming that $q_{+}$ and $q_{-}$ vary independently, the
integrability conditions are solved by any arbitrary function $\mathcal{V}=\mathcal{V}\left(q_{+},q_{-}\right)$
that fulfills 
\begin{equation}
\varphi_{\pm}=\frac{1}{2}\frac{\delta\mathcal{V}}{\delta q_{\mp}}\ .\label{eq:varphi +-}
\end{equation}
Hence, the surface integrals associated to the improved canonical
generators integrate as
\begin{equation}
\begin{aligned}Q_{\tilde{\mathcal{H}}}^{\pm}\left(T^{\pm}\right) & =\int d\phi T^{\pm}\mathcal{L}_{\pm}\ ,\end{aligned}
\label{Q H tilde+-}
\end{equation}
with 
\begin{equation}
\begin{aligned}\mathcal{L}_{\pm} & =\frac{1}{l\kappa}f_{\pm\pm}\mp\frac{lq_{\pm}}{8\pi^{2}}\left[q_{+}-q_{-}-\frac{2\pi}{l}\left(\frac{\delta\mathcal{V}}{\delta q_{-}}-\frac{\delta\mathcal{V}}{\delta q_{+}}\right)\right]+\frac{1}{4\pi}\mathcal{V}\ .\end{aligned}
\label{L+- integrado}
\end{equation}

Note that the surface integrals manifestly acquire contributions coming
from the electromagnetic field, and moreover, they also explicitly
depend on the arbitrary function $\mathcal{V}$ that has to be specified
by the boundary conditions in order to guarantee the integrability
of the global charges. This is a crucial point because otherwise,
if the variation of the canonical generators were not integrable,
the whole canonical structure would be spoiled, since it would conflict
with the fact that the Poisson brackets fulfill the Jacobi identity. 

\medskip{}

It must also be emphasized that fulfilling the integrability conditions
implies that only a subset of the purely geometric asymptotic symmetries
in \eqref{eq:Asymptotic Killing vectors} are reflected in the canonical
realization. Indeed, the amount of asymptotic symmetries that are
spanned by canonical generators depends on the choice of boundary
conditions specified by $\mathcal{V}$ . This is explained in the
next section.

\section{Compatibility of the boundary conditions with the asymptotic symmetries\label{sec:Compatibility}}

The boundary conditions are characterized by an arbitrary function
$\mathcal{V}$ that, according to \eqref{eq:varphi +-}, specifies
the functional relationship of the leading terms of the asymptotic
form of the electromagnetic gauge field, i.e., $\varphi_{\pm}=\varphi_{\pm}\left(q_{+},q_{-}\right)$.
Consequently, a generic choice of boundary conditions turns out to
be generically incompatible with the whole set of asymptotic symmetries
in \eqref{eq:Asymptotic Killing vectors}, because the transformations
rules of $\varphi_{\pm}$ and $q_{\pm}$ have to be consistent with
\begin{equation}
\begin{aligned}\delta\varphi_{\pm} & =\frac{\delta\varphi_{\pm}}{\delta q_{+}}\delta q_{+}+\frac{\delta\varphi_{\pm}}{\delta q_{-}}\delta q_{-}\;.\end{aligned}
\label{Variation of varphi+-}
\end{equation}
In what follows this is first analyzed for the asymptotic $U\left(1\right)$
gauge transformations and then for the asymptotic Killing vectors.

\subsection{Asymptotic $U\left(1\right)$ gauge transformations}

In the case of asymptotic $U\left(1\right)$ gauge transformations
spanned by \eqref{eq:Asymptotic Killing vectors} with $T^{\pm}=0$,
the functions $\varphi_{\pm}$ and $q_{\pm}$ transform according
to 
\begin{equation}
\delta\varphi_{\pm}=\partial_{\pm}\lambda\;;\;\delta q_{\pm}=0\;.\label{Asymptotic U(1)}
\end{equation}

Therefore, eq. \eqref{Variation of varphi+-} can only be compatible
with the transformation law in \eqref{Asymptotic U(1)} provided $\partial_{\pm}\lambda=0$. 

One then concludes that, if $q_{+}$ and $q_{-}$ are assumed to vary
independently, only the zero mode of $\lambda$ turns out to be consistent,
regardless the choice of boundary conditions. Hence, there is just
a single global charge that is conserved, which corresponds to \eqref{eq:electric charge}
evaluated for $\lambda=\lambda_{0}$ constant. This is the electric
charge $\mathcal{Q}$. 

Alternatively, the absence of additional conserved currents can be
easily verified from the conservation of the surface integral in \eqref{eq:electric charge}.
Indeed, making use of the leading term of the Maxwell equation in
the asymptotic region, given by $\partial_{+}q_{-}+\partial_{-}q_{+}=\partial_{\phi}q_{\phi}-l\partial_{t}q_{t}=0$,
requiring $\dot{Q}_{\mathcal{G}}\left(\lambda\right)=0$, implies
that $\lambda$ asymptotically approaches to a constant. One can then
get rid off the constant $U\left(1\right)$ parameter so that \eqref{eq:electric charge}
agrees with the Gauss formula.

It is worth pointing out that, as the remaining modes of $\lambda$
do not yield to conserved charges, they can not be regarded as legitimate
asymptotic symmetries, since they do not possess canonical generators\footnote{Note that additional $U\left(1\right)$ currents could be consistent,
but for different classes of boundary conditions that are not considered
in this work. Indeed, for instance, this would be the case if either
$q_{+}$ or $q_{-}$ is set to vanish without variation, so that the
modes of $Q_{\mathcal{G}}\left(\lambda\right)$ correspond to the
generators of left or right currents. }.

\subsection{Asymptotic Killing vectors }

Under the action of a generic asymptotic Killing vector spanned by
\eqref{eq:Asymptotic Killing vectors} with $\lambda=0$, the transformation
laws of $\varphi_{\pm}$ and $q_{\pm}$ are given by
\begin{equation}
\begin{aligned}\delta\varphi_{\pm} & =\partial_{\pm}\left(\varphi_{\pm}T^{\pm}\right)+T^{\mp}\partial_{\mp}\varphi_{\pm}+\frac{\ell}{4\pi}q_{\pm}\left(\partial_{+}T^{+}+\partial_{-}T^{-}\right)\ ,\\
\delta q_{\pm} & =\partial_{\pm}\left(q_{\pm}T^{\pm}\right)+T^{\mp}\partial_{\mp}q_{\pm}\ .
\end{aligned}
\label{Transformation laws varphi y q}
\end{equation}

In the case of (right) asymptotic symmetries spanned by $T^{+}$ (with
$T^{-}=0$), compatibility of the transformation law in \eqref{Transformation laws varphi y q}
with eq. \eqref{Variation of varphi+-} implies the following conditions:
\begin{equation}
\begin{aligned}\left(\frac{\ell}{4\pi}q_{+}+\frac{1}{2}\frac{\delta\mathcal{V}}{\delta q_{-}}-\frac{1}{2}q_{+}\frac{\delta^{2}\mathcal{V}}{\delta q_{+}\delta q_{-}}\right)\partial_{+}T^{+} & =0\;,\\
\left(\frac{\ell}{4\pi}q_{-}-\frac{1}{2}q_{+}\frac{\delta^{2}\mathcal{V}}{\delta q_{+}^{2}}\right)\partial_{+}T^{+} & =0\;.
\end{aligned}
\label{Preserving T+}
\end{equation}

Hence, for a generic choice of $\mathcal{V}$ , eqs. \eqref{Preserving T+}
can only be fulfilled provided $\partial_{+}T^{+}=0$ , i.e., just
for the zero mode of $T^{+}$. 

Requiring the whole set of modes of $T^{+}$ to be compatible with
the conditions in \eqref{Preserving T+} implies that the function
$\mathcal{V}$ has to satisfy both differential equations within the
round brackets in \eqref{Preserving T+}. Interestingly, up to an
additive constant, this singles out the set of boundary conditions
to be characterized by 
\begin{equation}
\begin{aligned}\mathcal{V} & =\mathcal{V}^{+}=\frac{lq_{+}q_{-}}{2\pi}\left[\log\left(q_{+}\right)-1\right]+q_{+}\mathcal{C}_{+}\left(q_{-}\right)\;,\end{aligned}
\label{NU+}
\end{equation}
where $\mathcal{C}_{+}$ is an arbitrary function of a single variable.

\medskip{}

If the asymptotic symmetries correspond to the ones spanned by $T^{-}$,
compatibility of \eqref{Transformation laws varphi y q} with \eqref{Variation of varphi+-}
implies that
\begin{equation}
\begin{aligned}\left(\frac{\ell}{4\pi}q_{-}+\frac{1}{2}\frac{\delta\mathcal{V}}{\delta q_{+}}-\frac{1}{2}q_{-}\frac{\delta^{2}\mathcal{V}}{\delta q_{+}\delta q_{-}}\right)\partial_{-}T^{-} & =0\;,\\
\left(\frac{\ell}{4\pi}q_{+}-\frac{1}{2}q_{-}\frac{\delta^{2}\mathcal{V}}{\delta q_{-}^{2}}\right)\partial_{-}T^{-} & =0\;.
\end{aligned}
\label{Preserving T-}
\end{equation}

Analogously, for generic boundary conditions specified by $\mathcal{V}$
, both eqs. in \eqref{Preserving T-} can only be satisfied for the
zero mode of $T^{-}$.

The family of boundary conditions that is consistent with the entire
set of modes of $T^{-}$ then corresponds to a choice of $\mathcal{V}$
that fulfills the differential equations in the round brackets of
\eqref{Preserving T-}. Up to an additive constant, the set is described
by an arbitrary function $\mathcal{C}_{-}\left(q_{+}\right)$ , so
that 
\begin{equation}
\begin{aligned}\mathcal{V} & \mathcal{=V}^{-}=\frac{lq_{+}q_{-}}{2\pi}\left[\log\left(q_{-}\right)-1\right]+q_{-}\mathcal{C}_{-}\left(q_{+}\right)\;.\end{aligned}
\label{NU -}
\end{equation}

\medskip{}

Remarkably, one then concludes that the whole set of asymptotic Killing
vectors, spanned by $T^{+}$ and $T^{-}$, turns out to be compatible
with the integrability of the canonical generators provided the boundary
conditions are specified by a very special set, being labeled by a
unique arbitrary constant parameter $\zeta$ without variation. This
is because simultaneously preserving left and right modes, implies
that $\mathcal{V}=\mathcal{V}^{+}=\mathcal{V}^{-}$. Hence, by virtue
of \eqref{NU+} and \eqref{NU -}, the very special set of boundary
conditions that preserves the full conformal group is determined by
the choice 
\begin{equation}
\mathcal{V}=\frac{l}{2\pi}q_{+}q_{-}\log\left(\zeta q_{+}q_{-}\right)\;.\label{NU magic}
\end{equation}

This result agrees with the one recently found in \cite{PRTT1}, which
was obtained from a completely different approach. Indeed, the same
choice of boundary conditions was recovered, but from requiring compatibility
with scaling and Lorentz symmetries of stationary spherically symmetric
solutions.

\subsection{Algebra of the canonical generators }

Following the results explained right above this section, one concludes
that the algebra of the canonical generators depends on the choice
of the boundary conditions specified by $\mathcal{V}$.

\medskip{}

For a generic choice of $\mathcal{V}$, the asymptotic symmetries
that are compatible with the boundary conditions just correspond to
the zero modes of $T^{+}$, $T^{-}$, $\lambda$ in \eqref{eq:Asymptotic Killing vectors}.
Hence, there are only three conserved charges given by \eqref{eq:electric charge}
and \eqref{Q H tilde+-} with constant parameters. Therefore, since
the asymptotic symmetries turn out to be spanned by $\partial_{t}$,
$\partial_{\phi}$ and a global $U\left(1\right)$ transformation,
the asymptotic symmetry group is given by $\mathbb{R}\otimes U\left(1\right)\otimes U\left(1\right)$,
so that the global charges correspond to the mass, the angular momentum
and the electric charge.

\medskip{}

In the case of the very special choice of boundary conditions given
by $\mathcal{V}$ in \eqref{NU magic}, the legitimate asymptotic
symmetries that possess well-defined canonical generators correspond
to the full conformal group spanned by $T^{+}$, $T^{-}$, and the
global $U\left(1\right)$ transformation generated by the zero mode
of $\lambda$. The surface integrals associated to the improved canonical
generators in \eqref{Q H tilde+-} are such that in this case eq.
\eqref{L+- integrado} reduces to

\begin{equation}
\begin{aligned}\mathcal{L}_{\pm} & =\frac{1}{l\kappa}f_{\pm\pm}+\frac{lq_{\pm}^{2}}{8\pi^{2}}\log\left(\zeta q_{+}q_{-}\right)\;.\end{aligned}
\label{eq: L+- final}
\end{equation}
It is worth highlighting that the $+r$ and $-r$ components of the
Einstein field equations, which generically read 
\begin{equation}
\begin{aligned}\partial_{\mp}f_{\pm\pm}+\frac{\kappa\ell^{2}}{8\pi^{2}}\left[q_{\mp}\partial_{\pm}q_{\pm}\pm\frac{2\pi}{\ell}q_{\pm}\left(\partial_{-}\left(\frac{\delta\mathcal{V}}{\delta q_{-}}\right)-\partial_{+}\left(\frac{\delta\mathcal{V}}{\delta q_{+}}\right)+\frac{3\ell}{2\pi}\partial_{+}q_{-}\right)\right]=0\ ,\end{aligned}
\label{Dynamical field eq}
\end{equation}
in this case just reduce to 
\begin{equation}
\partial_{\mp}\mathcal{L}_{\pm}=0\;.\label{eq:Conservacion de L+-}
\end{equation}
Therefore, since the Poisson brackets fulfill $\left\{ Q\left(Y_{1}\right),Q\left(Y_{2}\right)\right\} =\delta_{Y_{2}}Q\left(Y_{1}\right)$,
the algebra of the canonical generators can be readily obtained from
the transformation law of $q_{\pm}$ in \eqref{Transformation laws varphi y q},
\eqref{Asymptotic U(1)}, as well as from the one of $f_{\pm\pm}$,
which reads

\begin{equation}
\begin{aligned}\delta f_{\pm\pm} & =2f_{\pm\pm}\partial_{\pm}T^{\pm}+T^{\pm}\partial_{\pm}f_{\pm\pm}-\frac{\ell^{2}}{2}\partial_{\pm}^{3}T^{+}-\frac{\kappa l^{2}}{8\pi^{2}}q_{\pm}^{2}\left(\partial_{+}T^{+}+\partial_{-}T^{-}\right)+T^{\mp}\partial_{\mp}f_{\pm\pm}\ .\end{aligned}
\label{Delta f}
\end{equation}
Hence, the transformation laws of $\mathcal{L}_{\pm}$ in \eqref{eq: L+- final}
are given by
\begin{equation}
\delta\mathcal{L}_{\pm}=2\mathcal{L}_{\pm}\partial_{\pm}T^{\pm}+T^{\pm}\partial_{\pm}\mathcal{L}_{\pm}-\frac{l}{2\kappa}\partial_{\pm}^{3}T^{^{\pm}}\;.\label{eq:Ley de transformacion de L+-}
\end{equation}
Expanding in Fourier modes, $\mathcal{L_{\pm}}=\frac{1}{2\pi}\sum_{m}\mathcal{L}_{m}^{\pm}e^{im\phi}$,
the algebra of the canonical generators then reads $\mathcal{L}_{m}^{\pm}$
with the electric charge $\mathcal{Q}$ then reads 
\begin{equation}
\begin{aligned}i\left\{ \mathcal{L}_{m}^{\pm},\mathcal{L}_{n}^{\pm}\right\}  & =\left(m-n\right)\mathcal{L}_{m+n}^{\pm}+\frac{c}{12}m^{3}\delta_{m+n}\;,\\
i\left\{ \mathcal{L}_{m}^{\pm},\mathcal{Q}\right\}  & =0\;,\\
i\left\{ \mathcal{Q},\mathcal{Q}\right\}  & =0\;,
\end{aligned}
\label{Algebra canonical generators}
\end{equation}
with $c=3l/2G$, corresponding to the direct sum of $U\left(1\right)$
and two copies of the Virasoro algebra with the standard central extension.

\medskip{}

As a final remark of this section, it is worth mentioning that if
the boundary conditions were chosen so that $\mathcal{V=V}^{+}$ in
\eqref{NU+}, the legitimate asymptotic symmetries are spanned by
$T^{+}$, and the zero modes of $T^{-}$, $\lambda$. The canonical
generators then correspond to $\mathcal{L}_{m}^{+}$, $\mathcal{L}_{0}^{-}$,
and $\mathcal{Q}$, whose algebra is given by the direct sum of the
right copy of the Virasoro algebra with the Brown-Henneaux central
extension with two additional Abelian generators. Analogously, for
the choice $\mathcal{V=V}^{-}$ in \eqref{NU -}, the same occurs
for the left copy.

\section{Mass and angular momentum of electrically charged rotating black
holes\label{Sec. Black Hole}}

The rotating extension of the static BTZ black hole with electric
charge \cite{BTZ} was independently obtained in \cite{Clement-Q+J}
and \cite{MTZ} through different approaches. It is convenient to
express the solution as in \cite{PRTT1}, so that the line element
and the electromagnetic gauge field can be written as
\begin{equation}
\begin{aligned}ds^{2} & =-N^{2}F^{2}dt^{2}+\frac{d\rho^{2}}{F^{2}}+R^{2}\left(N^{\phi}dt+d\phi\right)^{2}\;,\\
A & =A_{t}dt+A_{\phi}d\phi\;,
\end{aligned}
\end{equation}
with
\begin{equation}
\begin{aligned}R^{2} & =\rho^{2}+\left(\frac{\omega^{2}}{1-\omega^{2}}\right)r_{+}^{2}+\frac{\kappa}{4\pi^{2}}\left(q_{t}\omega l\right)^{2}\log\left(\frac{\rho}{r_{+}}\right)\;,\\
N^{\phi} & =-\left(\frac{\omega}{1-\omega^{2}}\right)\left(\frac{\rho^{2}}{l^{2}}-F^{2}\right)\frac{l}{R^{2}}\;,\\
N^{2} & =\frac{\rho^{2}}{R^{2}}\;,\\
F^{2} & =\frac{\rho^{2}}{l^{2}}-\frac{r_{+}^{2}}{l^{2}}-\frac{\kappa}{4\pi^{2}}q_{t}^{2}\left(1-\omega^{2}\right)\log\left(\frac{\rho}{r_{+}}\right)\;,\\
A_{t} & =-\frac{q_{t}}{2\pi}\log\left(\frac{\rho}{l}\right)+\frac{\varphi_{t}}{l}\;,\\
A_{\phi} & =\frac{q_{t}\omega l}{2\pi}\log\left(\frac{\rho}{l}\right)+\varphi_{\phi}\;,
\end{aligned}
\end{equation}
where $r_{+}$, $\omega$, $q_{t}$, $\varphi_{t}$, $\varphi_{\phi}$
stand for arbitrary constants. Here, the Lagrange multipliers have
been chosen as $N_{\infty}=1$, and $N_{\infty}^{\phi}=\Phi=0$.

Note that in the asymptotic region, $\rho\rightarrow\infty$, the
metric components $g_{+-}$ and $g_{\rho\rho}$ possess terms that
behave as $\mathcal{O}\left(\log\left(\rho/l\right)\right)$ and $\mathcal{O}\left(\log\left(\rho/l\right)\rho^{-4}\right)$,
respectively; which as explained in section \eqref{sec:Geometric aspects},
can be consistently gauged away. Indeed, changing the coordinates
according to $x^{\pm}=\frac{t}{l}\pm\phi$, and $\rho=r+\frac{\kappa l^{2}}{16\pi^{2}}q_{t}^{2}\left(1-\omega^{2}\right)\log\left(\frac{r}{l}\right)r^{-1}$,
the asymptotic behaviour of the solution is given by

\begin{align}
g_{\pm\pm} & =\frac{\kappa l^{2}}{4\pi^{2}}q_{\pm}^{2}\log\left(\frac{r}{l}\right)+f_{\pm\pm}+\mathcal{O}\left(\log\left(\frac{r}{l}\right)r^{-1}\right)\ ,\nonumber \\
g_{+-} & =-\frac{r^{2}}{2}+f_{+-}+\mathcal{O}\left(\log\left(\frac{r}{l}\right)r^{-1}\right)\ ,\nonumber \\
g_{rr} & =\frac{l^{2}}{r^{2}}+\frac{f_{rr}}{r^{4}}+\mathcal{O}\left(\log\left(\frac{r}{l}\right)r^{-5}\right)\ ,\label{BlackHole metric and A}\\
g_{r\pm} & =0\ ,\nonumber \\
\nonumber \\
A_{\pm} & =-\frac{l}{2\pi}q_{\pm}\log\left(\frac{r}{l}\right)+\varphi_{\pm}+\mathcal{O}\left(\log\left(\frac{r}{l}\right)r^{-2}\right)\ ,\nonumber \\
A_{r} & =0\ ,\nonumber 
\end{align}
with
\begin{equation}
\begin{aligned}q_{\pm} & =\frac{1}{2}\left(q_{t}\pm q_{\phi}\right)\;;\;q_{\phi}=-\omega q_{t}\;,\\
f_{\pm\pm} & =\left(\frac{1\mp\omega}{1\pm\omega}\right)\frac{r_{+}^{2}}{4}-\frac{l^{2}q_{t}^{2}\kappa\left(1\mp\omega\right)^{2}}{16\pi^{2}}\log\left(\frac{r_{+}}{l}\right)\;,\\
f_{+-} & =\frac{4\pi^{2}r_{+}^{2}-l^{2}q_{t}^{2}\kappa\log\left(\frac{r_{+}}{l}\right)\left(1-\omega^{2}\right)}{16\pi^{2}}\;,\\
f_{rr} & =8l^{2}\pi^{2}r_{+}^{2}+l^{4}q_{t}^{2}\kappa\left(1-\omega^{2}\right)\left(1-\text{\ensuremath{\log}}\left(\frac{r_{+}^{2}}{l^{2}}\right)\right)\;,\\
\varphi_{\pm} & =\frac{1}{2}\left(\varphi_{t}\pm\varphi_{\phi}\right)\;,
\end{aligned}
\end{equation}
which perfectly fits within our asymptotic conditions in eqs. \eqref{eq:Asymptotic conditions metric},
\eqref{eq:Asymptotic conditions gauge field-2}. Therefore, the global
charges can be readily obtained from the surface integrals in \eqref{eq:electric charge}
and \eqref{Q H tilde+-}. 

\medskip{}

For a generic choice of boundary conditions, specified by an arbitrary
function $\mathcal{V}$ , the electric charge is given by $\mathcal{Q}=q_{t}$
, while the remaining ones are determined by $\begin{aligned}\mathcal{L}_{\pm}\end{aligned}
$ in eq. \eqref{L+- integrado}. Hence, the mass and the angular momentum
read 
\begin{equation}
\begin{aligned}M & =\frac{2\pi}{l}\left(\mathcal{L}_{+}+\mathcal{L}_{-}\right)\\
 & =\frac{\pi r_{+}^{2}}{\kappa l^{2}}\left(\frac{1+\omega^{2}}{1-\omega^{2}}\right)-\frac{q_{t}^{2}}{4\pi}\left[\omega^{2}+\left(1+\omega^{2}\right)\log\left(\frac{r_{+}}{l}\right)\right]+\frac{1}{l}\left(\mathcal{V}-q_{\phi}\frac{\partial\mathcal{V}}{\partial q_{\phi}}\right)\;,
\end{aligned}
\end{equation}
and 

\begin{equation}
\begin{aligned}J & =2\pi\left(\mathcal{L}_{+}-\mathcal{L}_{-}\right)\\
 & =-\frac{2\pi r_{+}^{2}\omega}{\kappa l\left(1-\omega^{2}\right)}+\frac{lq_{t}^{2}\omega}{4\pi}\left(1+\log\left(\frac{r_{+}^{2}}{l^{2}}\right)\right)-q_{t}\frac{\delta\mathcal{V}}{\delta q_{\phi}}\;,
\end{aligned}
\end{equation}
respectively, in full agreement with the expressions found in \cite{PRTT1}
from a minisuperspace of stationary spherically symmetric configurations\footnote{The difference in the sign of the angular momentum is because here
we have chosen the opposite orientation as compared with \cite{PRTT1}.}. In the case of boundary conditions with $\mathcal{V}=0$, the mass
and the angular momentum then reduce to the ones found in \cite{MTZ}.

\medskip{}

For the very special choice of boundary conditions that is compatible
with the full conformal symmetry at infinity, with $\mathcal{V}$
specified by \eqref{NU magic}, the zero modes of the Virasoro generators
are given by 
\begin{align}
\mathcal{L}_{\pm} & =\frac{r_{+}^{2}}{4l\kappa}\left(\frac{1\mp\omega}{1\pm\omega}\right)+\frac{lq_{t}^{2}\left(1\mp\omega\right)^{2}}{32\pi^{2}}\left[\log\left(\frac{\kappa}{8\pi^{2}}\frac{q_{t}^{2}l^{2}}{r_{+}^{2}}\left(1-\omega^{2}\right)\right)+\gamma-1\right]\;,
\end{align}
so that the mass and the angular momentum in this case read

\begin{equation}
\begin{aligned}M & =\frac{\pi}{\kappa}\left(\frac{1+\omega^{2}}{1-\omega^{2}}\right)\frac{r_{+}^{2}}{l^{2}}+\frac{q_{t}^{2}\left(1+\omega^{2}\right)}{8\pi}\left[\log\left(\frac{\kappa}{8\pi^{2}}\frac{q_{t}^{2}l^{2}}{r_{+}^{2}}\left(1-\omega^{2}\right)\right)+\gamma-1\right]\;,\end{aligned}
\end{equation}
and
\begin{equation}
J=-\frac{2l\omega}{1+\omega^{2}}M\;,
\end{equation}
 where the arbitrary fixed constant $\zeta$ that characterizes this
set of boundary conditions has been redefined as $\gamma=1+\log\left(\frac{2\pi^{2}}{\kappa}\zeta\right)$.
Therefore, according to \cite{PRTT1}, this parameter can be interpreted
as the slope that bounds the allowed region in the parameter space
where the solution describes a black hole. It is worth pointing out
that for this set of boundary conditions, the black hole energy spectrum
is nonnegative, and for a fixed value of the mass, the electric charge
possesses an upper bound.

\section{Ending remarks\label{Ending remarks}}

Improving the canonical generators as in \eqref{eq:H magico}, so
that they span the Lie derivative of the fields, was shown to be a
crucial point in order to unveil the asymptotic structure of electromagnetism
coupled to gravity with negative cosmological constant in three spacetime
dimensions. Despite the fall-off of the fields is extremely relaxed
as compared with the standard one, the global charges were shown to
be finite and manifestly acquire contributions from the electromagnetic
field. The existence of the canonical structure requires nontrivial
integrability conditions for the global charges to be fulfilled, which
implies a functional relationship between the leading terms in the
asymptotic form of the electromagnetic field. The global charges then
also explicitly depend on the choice of boundary conditions. This
effect has been also shown to occur for General Relativity coupled
to scalar and higher spin fields in three spacetime dimensions \cite{HMTZ},
\cite{HMTZ new reference}, \cite{Oscar}, \cite{PTT-HS}. 

In the case of the Maxwell field, the boundary conditions generically
break most of the purely geometric asymptotic symmetries, but nonetheless,
there is a very special choice that is compatible with the full conformal
symmetry at infinity\footnote{An intermediate situation is known to occur for asymptotically AdS$_{3}$
spacetimes in topologically massive gravity, where the integrability
conditions are even more stringent, so that the boundary conditions
preserve at most one chiral copy of the Virasoro algebra, while for
the other one only the zero mode survives \cite{HMT- More on...}.}. It is then clear that the possibility of studying holography along
the lines of the AdS$_{3}$/CFT$_{2}$ correspondence \cite{Maldacena},
\cite{Gubser}, \cite{Witten}, \cite{Kraus}, becomes widened to
the case of a fully backreacting Maxwell field. Moreover, the different
possibles choices of boundary conditions that are compatible with
the integrability of the global charges, might also be of interest
in the context of holographic superconductors \cite{Hartnoll-Herzog},
\cite{Maity}, \cite{Ren}, \cite{Faulkner-Iqbal}, \cite{Horowitz-Iqbal},
\cite{Chaturvedi-Sengupta}.

We would like to recall that our results have been obtained assuming
that the functions that characterize the leading terms in the fall-off
of the electromagnetic field, given by $q_{+}$ and $q_{-}$, are
assumed to vary independently. This has to be so in order to accommodate
the generic electrically charged rotating black hole solution within
the asymptotic conditions in \eqref{eq:Asymptotic conditions metric},
\eqref{eq:Asymptotic conditions gauge field-2}. Our set of asymptotically
AdS$_{3}$ conditions differs from the ones that have been previously
explored in the context of the Einstein-Maxwell theory \cite{Cadoni-Mass},
\cite{Glenn-1}. It is then worth pointing out that further inequivalent
sets of boundary conditions could also be constructed if one no longer
assumes that both $q_{+}$, $q_{-}$ are free to vary in an independent
way, so that the whole analysis would have to be done from scratch. 

As a final remark, it is worth mentioning that the effects of improving
the canonical generators according to \eqref{eq:H magico} do not
show up for asymptotically AdS spacetimes in $d\geq4$ dimensions.
Indeed, in the higher-dimensional case, the fall-off of the electromagnetic
field is slow enough so that the metric fulfills the standard asymptotically
AdS behaviour \cite{Henneaux-Teitelboim}, \cite{Henneaux}. Consequently,
since the time and angular components of the asymptotic Killing vectors
behave as $\mathcal{O}\left(1\right)$ in the asymptotic region, the
variation of the improvement term, given by $\int dS_{l}\xi^{\mu}A_{\mu}\delta p^{l}$,
necessarily vanishes when $r\rightarrow\infty$. Hence, the improved
and the standard generators of diffeomorphisms just differ in a term
that spans a proper gauge transformation which does not contribute
to the global charges.

\acknowledgments

We thank Glenn Barnich, Claudio Bunster, Oscar Fuentealba, Hernán
González, Marc Henneaux, Pu-Jian Mao, Cristián Martínez, Cedric Troessaert
and Jorge Zanelli for helpful comments. M.R. thanks Conicyt for financial
support. The work of D.T. is partially supported by the ERC Advanced
Grant ``SyDuGraM\textquotedblright , by a Marina Solvay fellowship,
by FNRS-Belgium (convention FRFC PDR T.1025.14 and convention IISN
4.4514.08) and by the ``Communauté Française de Belgique\textquotedblright{}
through the ARC program. This research has been partially supported
by Fondecyt grants Nº 11130262, 11130260, 1130658, 1121031. Centro
de Estudios Científicos (CECs) is funded by the Chilean Government
through the Centers of Excellence Base Financing Program of Conicyt.


\begin{thebibliography}{10}
\bibitem{Reissner}H.~Reissner,   ``Über die Eigengravitation des elektrischen Feldes nach der Einsteinschen Theorie,''   Annalen der Physik (in German) 50: 106-120  (1916).

\bibitem{Nordstrom}G.~Nordström,   ``On the energy of the Gravitational Field in Einstein's Theory,''    Verhandl. Koninkl. Ned. Akad. Wetenschap., Afdel. Natuurk., Amsterdam 26: 1201-1208  (1918).

\bibitem{Hawking-Ellis} S.~W.~Hawking and G.~F.~R.~Ellis,   ``The Large Scale Structure of Space-Time''.   

\bibitem{Townsed}P.~K.~Townsend,   ``Black holes: Lecture notes,''   gr-qc/9707012.   

\bibitem{Romans}L.~J.~Romans,   ``Supersymmetric, cold and lukewarm black holes in cosmological Einstein-Maxwell theory,''   Nucl.\ Phys.\ B {\bf 383}, 395 (1992)   doi:10.1016/0550-3213(92)90684-4   [hep-th/9203018].   

\bibitem{Kerr}R.~P.~Kerr,   ``Gravitational field of a spinning mass as an example of algebraically special metrics,''   Phys.\ Rev.\ Lett.\  {\bf 11}, 237 (1963).   doi:10.1103/PhysRevLett.11.237   

\bibitem{Newman}E.~T.~Newman, R.~Couch, K.~Chinnapared, A.~Exton, A.~Prakash and R.~Torrence,   ``Metric of a Rotating, Charged Mass,''   J.\ Math.\ Phys.\  {\bf 6}, 918 (1965).   doi:10.1063/1.1704351   

\bibitem{BTZ} M.~Banados, C.~Teitelboim and J.~Zanelli,   ``The Black hole in three-dimensional space-time,''   Phys.\ Rev.\ Lett.\  {\bf 69}, 1849 (1992)   doi:10.1103/PhysRevLett.69.1849   [hep-th/9204099].   

\bibitem{Clement-Q+J} G.~Clement,   ``Classical solutions in three-dimensional Einstein-Maxwell cosmological gravity,''   Class.\ Quant.\ Grav.\  {\bf 10}, L49 (1993).   doi:10.1088/0264-9381/10/5/002   

\bibitem{MTZ}C.~Martinez, C.~Teitelboim and J.~Zanelli,   ``Charged rotating black hole in three space-time dimensions,''   Phys.\ Rev.\ D {\bf 61}, 104013 (2000)   doi:10.1103/PhysRevD.61.104013   [hep-th/9912259].   

\bibitem{Deser-Mazur} S.~Deser and P.~O.~Mazur,   ``Static Solutions in $D=3$ Einstein-maxwell Theory,''   Class.\ Quant.\ Grav.\  {\bf 2}, L51 (1985).   doi:10.1088/0264-9381/2/3/003   

\bibitem{PRTT1} A.~Perez, M.~Riquelme, D.~Tempo and R.~Troncoso,   ``Conserved charges and black holes in the Einstein-Maxwell theory on AdS$_{3}$ reconsidered,''   JHEP {\bf 1510}, 161 (2015)   doi:10.1007/JHEP10(2015)161   [arXiv:1509.01750 [hep-th]].   

\bibitem{ClementSpin} G.~Clement,   ``Spinning charged BTZ black holes and selfdual particle - like solutions,''   Phys.\ Lett.\ B {\bf 367}, 70 (1996)   doi:10.1016/0370-2693(95)01464-0   [gr-qc/9510025].   

\bibitem{Chan - Comment}K.~C.~K.~Chan,   ``Comment on the calculation of the angular momentum for the (anti)selfdual charged spinning BTZ black hole,''   Phys.\ Lett.\ B {\bf 373}, 296 (1996)   doi:10.1016/0370-2693(96)00145-1   [gr-qc/9509032].   

\bibitem{Kamata Koikawa mass}M.~Kamata and T.~Koikawa,   ``(2+1)-dimensional charged black hole with (anti-)selfdual Maxwell fields,''   Phys.\ Lett.\ B {\bf 391}, 87 (1997)   doi:10.1016/S0370-2693(96)01461-X   [hep-th/9605114].   

\bibitem{Dias-Lemos} O.~J.~C.~Dias and J.~P.~S.~Lemos,   ``Rotating magnetic solution in three-dimensional Einstein gravity,''   JHEP {\bf 0201}, 006 (2002)   doi:10.1088/1126-6708/2002/01/006   [hep-th/0201058].   

\bibitem{Clement-Mass} G.~Clement,   ``Black hole mass and angular momentum in 2+1 gravity,''   Phys.\ Rev.\ D {\bf 68}, 024032 (2003)   doi:10.1103/PhysRevD.68.024032   [gr-qc/0301129].   

\bibitem{Cadoni-Mass}M.~Cadoni, M.~Melis and M.~R.~Setare,   ``Microscopic entropy of the charged BTZ black hole,''   Class.\ Quant.\ Grav.\  {\bf 25}, 195022 (2008)   doi:10.1088/0264-9381/25/19/195022   [arXiv:0710.3009 [hep-th]].   

\bibitem{Myung-Mass}Y.~S.~Myung, Y.~W.~Kim and Y.~J.~Park,   ``Entropy function approach to charged BTZ black hole,''   Gen.\ Rel.\ Grav.\  {\bf 42}, 1919 (2010)   doi:10.1007/s10714-010-0969-5   [arXiv:0903.2109 [hep-th]].   

\bibitem{Jensen} K.~Jensen,   ``Chiral anomalies and AdS/CMT in two dimensions,''   JHEP {\bf 1101}, 109 (2011)   doi:10.1007/JHEP01(2011)109   [arXiv:1012.4831 [hep-th]].   

\bibitem{Garcia-review-mass} A.~A.~Garcia-Diaz,   ``Three dimensional stationary cyclic symmetric Einstein-Maxwell solutions; energy, mass, momentum, and algebraic tensors characteristics,''   arXiv:1307.6652 [gr-qc].   

\bibitem{Hendi-Mass} S.~H.~Hendi, S.~Panahiyan and R.~Mamasani,   ``Thermodynamic stability of charged BTZ black holes: Ensemble dependency problem and its solution,''   Gen.\ Rel.\ Grav.\  {\bf 47}, no. 8, 91 (2015)   doi:10.1007/s10714-015-1932-2   [arXiv:1507.08496 [gr-qc]].   

\bibitem{Frassino =000026 Mureika}A.~M.~Frassino, R.~B.~Mann and J.~R.~Mureika,   ``Lower-Dimensional Black Hole Chemistry,''   arXiv:1509.05481 [gr-qc].   

\bibitem{Brown-Henne}J.~D.~Brown and M.~Henneaux,   ``Central Charges in the Canonical Realization of Asymptotic Symmetries: An Example from Three-Dimensional Gravity,''   Commun.\ Math.\ Phys.\  {\bf 104}, 207 (1986).   doi:10.1007/BF01211590   

\bibitem{Henne-Teitel AdS}M.~Henneaux and C.~Teitelboim,   ``Asymptotically anti-De Sitter Spaces,''   Commun.\ Math.\ Phys.\  {\bf 98}, 391 (1985).   doi:10.1007/BF01205790   

\bibitem{HMTZ-D}M.~Henneaux, C.~Martinez, R.~Troncoso and J.~Zanelli,   ``Asymptotic behavior and Hamiltonian analysis of anti-de Sitter gravity coupled to scalar fields,''   Annals Phys.\  {\bf 322}, 824 (2007)   doi:10.1016/j.aop.2006.05.002   [hep-th/0603185].   

\bibitem{HMT-Topologically}M.~Henneaux, C.~Martinez and R.~Troncoso,   ``Asymptotically anti-de Sitter spacetimes in topologically massive gravity,''   Phys.\ Rev.\ D {\bf 79}, 081502 (2009)   doi:10.1103/PhysRevD.79.081502   [arXiv:0901.2874 [hep-th]].   

\bibitem{HMT- More on...}M.~Henneaux, C.~Martinez and R.~Troncoso,   ``More on Asymptotically Anti-de Sitter Spaces in Topologically Massive Gravity,''   Phys.\ Rev.\ D {\bf 82}, 064038 (2010)   doi:10.1103/PhysRevD.82.064038   [arXiv:1006.0273 [hep-th]].   

\bibitem{KK} M.~Kamata and T.~Koikawa,   ``The Electrically charged BTZ black hole with self (antiself) dual Maxwell field,''   Phys.\ Lett.\ B {\bf 353}, 196 (1995)   doi:10.1016/0370-2693(95)00583-7   [hep-th/9505037].   

\bibitem{Hirschmann-Welch} E.~W.~Hirschmann and D.~L.~Welch,   ``Magnetic solutions to (2+1) gravity,''   Phys.\ Rev.\ D {\bf 53}, 5579 (1996)   doi:10.1103/PhysRevD.53.5579   [hep-th/9510181].   

\bibitem{Cataldo-Salgado} M.~Cataldo and P.~Salgado,   ``Static Einstein-Maxwell solutions in (2+1)-dimensions,''   Phys.\ Rev.\ D {\bf 54}, 2971 (1996).   doi:10.1103/PhysRevD.54.2971   

\bibitem{Kamata-Koikawa-mass} M.~Kamata and T.~Koikawa,   ``(2+1)-dimensional charged black hole with (anti-)selfdual Maxwell fields,''   Phys.\ Lett.\ B {\bf 391}, 87 (1997)   doi:10.1016/S0370-2693(96)01461-X   [hep-th/9605114].   

\bibitem{Cat-Salgado}M.~Cataldo and P.~Salgado,   ``Three dimensional extreme black hole with self (anti-self) dual Maxwell field,''   Phys.\ Lett.\ B {\bf 448}, 20 (1999).   doi:10.1016/S0370-2693(99)00035-0   

\bibitem{Cataldo}M.~Cataldo,   ``Azimuthal electric field in a static rotationally symmetric (2+1)-dimensional space-time,''   Phys.\ Lett.\ B {\bf 529}, 143 (2002)   doi:10.1016/S0370-2693(02)01188-7   [gr-qc/0201047].   

\bibitem{crisostomo}M.~Cataldo, J.~Crisostomo, S.~del Campo and P.~Salgado,   ``On magnetic solution to (2+1) Einstein-Maxwell gravity,''   Phys.\ Lett.\ B {\bf 584}, 123 (2004)   doi:10.1016/j.physletb.2004.01.062   [hep-th/0401189].   

\bibitem{matyjasek}J.~Matyjasek and O.~B.~Zaslavskii,   ``Extremal limit for charged and rotating (2+1)-dimensional black holes and Bertotti-Robinson geometry,''   Class.\ Quant.\ Grav.\  {\bf 21}, 4283 (2004)   doi:10.1088/0264-9381/21/17/014   [gr-qc/0404090].   

\bibitem{Garcia annals}A.~A.~Garcia-Diaz,   ``Three dimensional stationary cyclic symmetric Einstein-Maxwell solutions; black holes,''   Annals of Physics {\bf {\bf 324}}, 2004 (2009)   doi:10.1016/j.aop.2009.04.004   [arXiv:1307.6655 [gr-qc]].   

\bibitem{ADM}R.~L.~Arnowitt, S.~Deser and C.~W.~Misner,   ``The Dynamics of general relativity,''   Gen.\ Rel.\ Grav.\  {\bf 40}, 1997 (2008)   doi:10.1007/s10714-008-0661-1   [gr-qc/0405109].   

\bibitem{Hanson Regge Teitelboim} A.~J.~Hanson, T.~Regge and C.~Teitelboim,   ``Constrained Hamiltonian Systems,''   RX-748, PRINT-75-0141 (IAS,PRINCETON).   

\bibitem{Henneaux-Teitelboim  Z} M.~Henneaux and C.~Teitelboim,   ``The Cosmological Constant As A Canonical Variable,''   Phys.\ Lett.\ B {\bf 143}, 415 (1984).   doi:10.1016/0370-2693(84)91493-X   

\bibitem{Bunster-Perez} C.~Bunster and A.~Pérez,   ``Superselection rule for the cosmological constant in three-dimensional spacetime,''   Phys.\ Rev.\ D {\bf 91}, no. 2, 024029 (2015)   doi:10.1103/PhysRevD.91.024029   [arXiv:1412.1492 [hep-th]].   

\bibitem{Regge-Teitelboim}T.~Regge and C.~Teitelboim,   ``Role of Surface Integrals in the Hamiltonian Formulation of General Relativity,''   Annals Phys.\  {\bf 88}, 286 (1974).   doi:10.1016/0003-4916(74)90404-7   

\bibitem{HMTZ} M.~Henneaux, C.~Martinez, R.~Troncoso and J.~Zanelli,   ``Black holes and asymptotics of 2+1 gravity coupled to a scalar field,''   Phys.\ Rev.\ D {\bf 65}, 104007 (2002)   doi:10.1103/PhysRevD.65.104007   [hep-th/0201170].   

\bibitem{HMTZ new reference}M.~Henneaux, C.~Martinez, R.~Troncoso and J.~Zanelli,   ``Asymptotically anti-de Sitter spacetimes and scalar fields with a logarithmic branch,''   Phys.\ Rev.\ D {\bf 70}, 044034 (2004)   doi:10.1103/PhysRevD.70.044034   [hep-th/0404236].   

\bibitem{Oscar}M.~Cardenas, O.~Fuentealba and C.~Martínez,   ``Three-dimensional black holes with conformally coupled scalar and gauge fields,''   Phys.\ Rev.\ D {\bf 90}, no. 12, 124072 (2014)   doi:10.1103/PhysRevD.90.124072   [arXiv:1408.1401 [hep-th]].   

\bibitem{PTT-HS}A.~Perez, D.~Tempo and R.~Troncoso,   ``Higher spin gravity in 3D: Black holes, global charges and thermodynamics,''   Phys.\ Lett.\ B {\bf 726}, 444 (2013)   doi:10.1016/j.physletb.2013.08.038   [arXiv:1207.2844 [hep-th]].   

\bibitem{Maldacena}J.~M.~Maldacena,   ``The Large N limit of superconformal field theories and supergravity,''   Int.\ J.\ Theor.\ Phys.\  {\bf 38}, 1113 (1999)   [Adv.\ Theor.\ Math.\ Phys.\  {\bf 2}, 231 (1998)]   doi:10.1023/A:1026654312961   [hep-th/9711200].   

\bibitem{Gubser}S.~S.~Gubser, I.~R.~Klebanov and A.~M.~Polyakov,   ``Gauge theory correlators from noncritical string theory,''   Phys.\ Lett.\ B {\bf 428}, 105 (1998)   doi:10.1016/S0370-2693(98)00377-3   [hep-th/9802109].   

\bibitem{Witten}E.~Witten,   ``Anti-de Sitter space and holography,''   Adv.\ Theor.\ Math.\ Phys.\  {\bf 2}, 253 (1998)   [hep-th/9802150].   

\bibitem{Kraus} P.~Kraus,   ``Lectures on black holes and the AdS(3) / CFT(2) correspondence,''   Lect.\ Notes Phys.\  {\bf 755}, 193 (2008)   [hep-th/0609074].   

\bibitem{Hartnoll-Herzog} S.~A.~Hartnoll, C.~P.~Herzog and G.~T.~Horowitz,   ``Building a Holographic Superconductor,''   Phys.\ Rev.\ Lett.\  {\bf 101}, 031601 (2008)   doi:10.1103/PhysRevLett.101.031601   [arXiv:0803.3295 [hep-th]].   

\bibitem{Maity}D.~Maity, S.~Sarkar, N.~Sircar, B.~Sathiapalan and R.~Shankar,   ``Properties of CFTs dual to Charged BTZ black-hole,''   Nucl.\ Phys.\ B {\bf 839}, 526 (2010)   doi:10.1016/j.nuclphysb.2010.06.012   [arXiv:0909.4051 [hep-th]].   

\bibitem{Ren}J.~Ren,   ``One-dimensional holographic superconductor from AdS$_3$/CFT$_2$ correspondence,''   JHEP {\bf 1011}, 055 (2010)   doi:10.1007/JHEP11(2010)055   [arXiv:1008.3904 [hep-th]].   

\bibitem{Faulkner-Iqbal} T.~Faulkner and N.~Iqbal,   ``Friedel oscillations and horizon charge in 1D holographic liquids,''   JHEP {\bf 1307}, 060 (2013)   doi:10.1007/JHEP07(2013)060   [arXiv:1207.4208 [hep-th]].   

\bibitem{Horowitz-Iqbal} G.~T.~Horowitz, N.~Iqbal and J.~E.~Santos,   ``Simple holographic model of nonlinear conductivity,''   Phys.\ Rev.\ D {\bf 88}, no. 12, 126002 (2013)   doi:10.1103/PhysRevD.88.126002   [arXiv:1309.5088 [hep-th]].   

\bibitem{Chaturvedi-Sengupta} P.~Chaturvedi and G.~Sengupta,   ``Rotating BTZ Black Holes and One Dimensional Holographic Superconductors,''   Phys.\ Rev.\ D {\bf 90}, no. 4, 046002 (2014)   doi:10.1103/PhysRevD.90.046002   [arXiv:1310.5128 [hep-th]].   

\bibitem{Glenn-1}G.~Barnich and P.~H.~Lambert,   ``Einstein-Yang-Mills theory: Asymptotic symmetries,''   Phys.\ Rev.\ D {\bf 88}, 103006 (2013)   doi:10.1103/PhysRevD.88.103006   [arXiv:1310.2698 [hep-th]].   

\bibitem{Henneaux-Teitelboim} M.~Henneaux and C.~Teitelboim,   ``Asymptotically anti-De Sitter Spaces,''   Commun.\ Math.\ Phys.\  {\bf 98}, 391 (1985).   doi:10.1007/BF01205790   

\bibitem{Henneaux}M.~Henneaux,   ``Asymptotically Anti-de Sitter Universes In D = 3, 4 And Higher Dimensions,'' in Proceedings of the Fourth Marcel Grossmann Meeting on General Relativity, R. Rufini (ed.). (Elsevier Science Publishiers B. V., 1986).   
\end{thebibliography}
\end{document}